\documentstyle[preprint,aps,epsf]{revtex}

\def\c12{$^{12}_\Lambda$C}
\def\b11{$^{11}_\Lambda$B}
\def\he5{$^{5}_\Lambda$He}
\def\LN{\Lambda N }

\newcommand{\sst}[1]{\scriptscriptstyle{#1}}

\begin{document}
\draft
%\preprint{WU93-001}
\title{
The weak strangeness production reaction pn $\to$ p$\Lambda$ \\ 
in a one-boson-exchange model}

\author{A. Parre\~no \footnote{Present address: INT, University of Washington, 
Seattle, WA 98195}, A. Ramos}
\address{
Departament d'Estructura i Constituents de la Mat\`eria, Universitat de
Barcelona,\\
Diagonal 647, 08028 Barcelona, Spain}
\author{N.G. Kelkar}
\address{Nuclear Physics Division, Bhabha Atomic Research Centre, Trombay,
 Mumbai-400 085, India}
\author{C. Bennhold}
\address{
Center of Nuclear Studies, Department of Physics,
The George Washington University, Washington, DC 20052, USA}
%\date{\today}
\maketitle

\begin{abstract}
The weak production of Lambdas in nucleon-nucleon scattering is 
studied in a meson-exchange framework.  The weak transition operator
for the $NN \to N \Lambda$ reaction is identical to a previously developed
weak strangeness-changing transition potential $\Lambda N \to NN$ that 
describes the nonmesonic decay of hypernuclei.  The initial $NN$ and final $YN$
state interaction has been included by using realistic baryon-baryon 
forces that describe the available elastic scattering data.  The total and
differential cross sections as well as the parity-violating asymmetry are
studied for the reaction pn $\to$ p$\Lambda$. These observables are found
to be sensitive to the opening of the $\Sigma$ production channel,
the choice of the strong interaction potential and the structure of
the weak transition potential.
\end{abstract}

\section{Introduction}
\label{intro}
Over the last several decades, the Standard Model of weak interactions 
has been thoroughly tested by a vast amount of data for leptonic and semi-
leptonic decays and reactions.  Hadronic weak interactions are in general
more difficult to study experimentally since they are usually obscured
by the presence of the much larger strong interaction.  
This requires employing processes in which the strong force
cannot participate due to overriding symmetry principles.  In the case of the 
weak nucleon-nucleon interaction it was realized more than 40 years 
ago\cite{feynman58} that 
the current-current form of the weak interaction dictates the presence
of a weak transition between nucleons which would lead to parity impurities
in nuclear states which are of first order in the weak coupling.  Using
these parity nonconserving observables in many experiments on nuclear gamma
and alpha transitions, polarized $NN$ scattering, as well as the recent
first measurement of the nuclear anapole moment\cite{wood97}, much has been 
learned about the weak nucleon-nucleon interaction\cite{haxton85}.

The situation is very different for the flavor-changing 
baryon-baryon interaction.  Soon after the discovery of hypernuclei
it was recognized that the $\Lambda$ inside the nuclear medium
decays not through its Pauli-blocked mesonic decay channel, but
predominantly through the $\Delta S$ =1 nonmesonic transition
$\Lambda N \to N N$, thus opening a door to the study of the weak
strangeness-changing hyperon-nucleon force.  However, experimental
progress in this field of weak hypernuclear decays has been slow
until recently due to the difficult multi-coincidence, low count-rate
nature of these measurements.  In recent years the situation has improved
significantly due to a series of new experiments at BNL and KEK.
On the theoretical side it became clear that in order to 
unambiguously extract the weak $\Lambda N \to N N$ transition potential 
significant effort must be spent to account for the nuclear structure
effects as accurately as possible.  When this task was recently
completed\cite{PRB97} it became apparent that major discrepancies
between theory and experiment cannot be due to the
underlying nuclear structure but have to arise from the nature
of the weak transition potential itself. 

Even with the nuclear structure input under control it has
become desirable to measure the process $NN \to N \Lambda$
directly since the hypernuclear decay can only probe the
reaction at one well-defined kinematic setting.  Though considered
impossible for many years since this process is reduced
in cross section by around 12 orders of magnitude from 
the standard elastic $NN$ scattering, recent progress
in experimental and accelerator technology may have
brought measuring this process within reach \cite{kishi95}. 
It is therefore
timely to provide predictions of various observables 
based on the transition potential used in the nonmesonic
hypernuclear decay.

In Sec.~\ref{dcs} of this paper, we present the expressions for
the matrix elements and the cross section. Sec.~\ref{wkp} 
briefly describes the transition operator derived in
Ref.\cite{PRB97}. Our results are discussed in Sec.~\ref{res} 
and summarized in Sec.~\ref{sum}.

\section{Differential Cross-Section}
\label{dcs}

\begin{figure}[hbt]
\caption{Feynman diagram for the np $\to \Lambda$p reaction}
\vspace*{0.5cm}
       \setlength{\unitlength}{1mm}
       \begin{picture}(100,180)
       \put(-20,-60){\epsfxsize=19cm \epsfbox{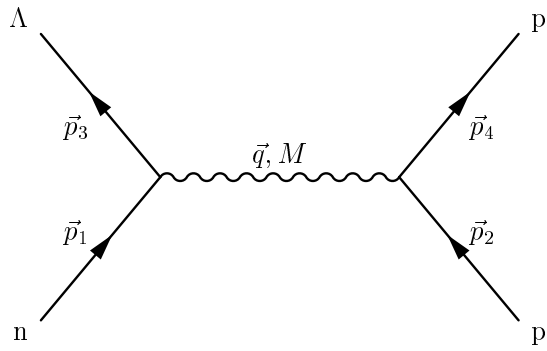}}
       \end{picture}
\label{fig:diag1}
\end{figure}

The differential cross section per unit solid angle in the center-of-mass
system for the reaction $p n \to p \Lambda$ as depicted in Fig.~\ref{fig:diag1} 
is given by the expression

\begin{equation}
\frac{d {\sigma}}{d \Omega} = (2 \pi)^4 \frac{1}{s} 
\frac{|\vec p_F|}{|\vec p_I|} E_1 E_2 E_3 E_4 
\frac{1}{(2s_1+1)(2s_2+1)}
\sum_{m_{s1}} \sum_{m_{s2}} \sum_{m_{s3}} \sum_{m_{s4}}
\mid {\cal M}_{FI} \mid^2  \,\, ,
\label{eq:cs2}
\end{equation}
with $\sqrt{s}=E_1 + E_2=E_3 + E_4$ the total available energy in the 
center-of-mass system and
${\vec p}_I$ and ${\vec p}_F$ the relative momenta of the particles
in the initial and final states respectively. 
In a plane wave Born approximation (PWBA) the weak transition matrix 
elements read

\begin{equation}
{\cal M}_{FI} = \langle {\vec p}_3 m_{s_3} t_3, {\vec p}_4 m_{s_4} t_4
| V^{w} |
\overline {{\vec p}_1 m_{s_1} t_1, {\vec p}_2 m_{s_2} t_2 \rangle}
\label{eq:cs3}
\end{equation}
where the overline stands for the antisymmetric combination of the two
particle states $\{1\}$ and $\{2\}$ and $V^{w}$ is the 
nonrelativistic weak transition potential.

Accounting for the
locality of the weak potential, the direct term of the previous matrix 
elements can be written in the distorted wave Born approximation (DWBA) 
as

\begin{eqnarray}
{\cal M}_{FI}&=& \sum_{S_F M_{S_F}} \sum_{S_I M_{S_I}} \sum_{T M_T} 
\langle \frac{1}{2} m_{s_3} \frac{1}{2} m_{s_4} | S_F M_{S_F} \rangle
\langle \frac{1}{2} t_3 \frac{1}{2} t_4 | T M_T  \rangle \nonumber \\
&\times& 
\langle \frac{1}{2} m_{s_1} \frac{1}{2} m_{s_2} | S_I M_{S_I} \rangle
\langle \frac{1}{2} t_1 \frac{1}{2} t_2 | T M_T  \rangle \nonumber \\
&\times& 
\int d\Omega \int r^2 dr \,[\Psi_{\Lambda N}^{(-)}({\vec p}_F,{\vec r})]^* 
{\chi^\dagger}^{T}_{M_T} V^{w}({\vec r}\,) 
\Psi_{NN}^{(+)}({\vec p}_I,{\vec r}) \chi^{T}_{M_T} 
%\left(\frac{1-(-1)^{L_I+S_I+T}}{\sqrt{2}}\right) 
\end{eqnarray}
where $ V^{w}({\vec r}\,)$ contains the radial, angular and isospin 
dependence of the weak transition
potential and $ \Psi_{\Lambda N}^{(-)}$ 
($\Psi_{NN}^{(+)}$) stands
for the distorted $\LN$ ($NN$) wave function. 

In Sec.~\ref{wkp} it is shown how the weak potential can be decomposed as

\begin{equation}    
V^{w}({\vec r}\,) = \sum_{i} \sum_{\alpha} V_{\alpha}^{(i)}({\vec r}\,) =
\sum_{i} \sum_{\alpha} V_{\alpha}^{(i)}(r) V_{\alpha}^{(i)}({\hat r}) 
{\hat I}_{\alpha}^{(i)} \ ,
\end{equation}
where the index i sums over mesons and $\alpha$ over the different spin 
channels. The radial part of the potential is denoted by 
$V_{\alpha}^{(i)}(r)$, the piece containing the angular and spin dependence
by $V_{\alpha}^{(i)}({\hat r})$ and ${\hat I}_{\alpha}^{(i)}$ denotes the
appropriate isospin operator for each meson. 

Using the partial wave decomposition for the distorted waves, working in
the coupled basis formalism, $LS(J)$ (see appendix), 
and assuming ${\hat p_I}$ parallel to the z-axis, the modulus squared of the
weak matrix elements for the $p n \to p \Lambda$ reaction finally reads

\begin{eqnarray}
|{\cal M}_{FI}|^2&=& \sum_{S_F M_{S_F}} \,\sum_{S_I M_{S_I}}\, \sum_T
\,\,\langle \frac{1}{2} t_1 \frac{1}{2} t_2 | T 0 \rangle ^2
\,\,\langle T 0 | \frac{1}{2} t_3 \frac{1}{2} t_4 \rangle ^2
\nonumber \\
&\times& \left | \frac{2}{\pi} \sum_{i} \sum_{\alpha} 
\langle T 0 |I_{\alpha}^{(i)} | T 0 \rangle
\,\sum_J \, \sum_{L_F' S_F'}\, \sum_{L_F}\, \sum_{L_I}\, \sum_{L_I'}
\,i^{(L_I' - L_F')} \right. 
\nonumber \\
&\times& \,\langle L_F M_{L_F} S_F M_{S_F} | J M \rangle
\,Y_{L_F M_{L_F}} ({\hat p}_F) 
\langle L_I M_{L_I} S_I M_{S_I} | J M_{S_I} \rangle 
\,\sqrt{\frac{2 L_I + 1}{4 \pi}}
\nonumber \\
&\times& \int r^2 dr 
\,[\psi^{(-) *}_{\Lambda N}]^J_{L_F' S_F', L_F S_F} (k_F, r)
\,V_{\alpha}^{(i)}(r)
\,[\psi^{(+)}_{NN}]^J_{L_I' S_I, L_I S_I} (k_I, r)
\nonumber \\
&\times& \left. \int d\Omega 
\,{{\cal J}^{\dagger}}^{J M}_{L_F' S_F'} ({\hat r})
\,V_{\alpha}^{(i)}({\hat r}) 
\,{\cal J}^{J M}_{L_I' S_I'} ({\hat r})
\,\,\left(\frac{1-(-1)^{L_I'+S_I+T}}{\sqrt{2}} \right) \right|^2 .
\label{eq:mef}
\end{eqnarray}

The distorted radial wave functions in the above equation are generated from 
a T-matrix which is constructed using the nucleon-nucleon ($NN$) and 
hyperon-nucleon ($YN$) strong potentials. We make use of the Nijmegen 93 
\cite{nijmnn} and 
Bonn B \cite{bonn} $NN$ potentials and the Nijmegen soft-core \cite{nijmln} 
and J\"ulich
\cite{HHS89} $YN$ potentials. Comparison between the results obtained using the
 different interaction 
models is made in Sec.~\ref{res}.

\section{The weak potential} 
\label{wkp}

In Ref. \cite{PRB97} a one-boson-exchange model is developed to describe the
$\Lambda N \to NN$ transition, where the pseudoscalar $\pi,\eta$,K and
vector $\rho,\omega$ and K$^*$ mesons mediate the interaction.
We use this model in order 
to describe the present inverse reaction.
In this study we have refrained from considering the transition
$p n \to N \Sigma \to p \Lambda$, since it was found to be an order of
magnitude smaller than that for the direct $\Lambda$ production \cite{HHKST95}.

The nonrelativistic reduction of the Feynman amplitude, corresponding
to the diagram depicted in Fig.~\ref{fig:diag1}, leads to the 
nonrelativistic weak potential in momentum space, which for the exchange
of pseudoscalar mesons takes the form

\begin{equation}
V_{ps}^{w} ({\vec q}\,) = - G_F m_\pi^2
\frac{g}{2M_N} \left(
{\hat A} + \frac{{\hat B}}{2\overline{M}}
%\mbox{\boldmath $\sigma$}_1 {\bf q} \right)
\mbox{${\vec \sigma}$}_1 {\vec q}\, \right)
%\frac{\mbox{\boldmath $\sigma$}_2 {\bf q} }{{\bf q}^2+\mu^2} \, \ ,
\frac{\mbox{$\vec{\sigma}$}_2 {\vec q} }{{\vec q\,}^2+\mu^2} \, \ ,
\label{eq:pspot}
\end{equation}
where $G_F m_\pi^2 = 2.21 \times 10^{-7}$ is the Fermi coupling constant, 
${\vec q}$ is the momentum carried by the meson (M) directed towards the
strong vertex, $g=g_{\rm {\scriptscriptstyle NNM}}$
the strong coupling constant for the NNM vertex, $\mu$
the meson mass,
$M_N$ the nucleon mass and $\overline M$
the average between the nucleon and $\Lambda$ masses.
The operators
${\hat A}$ and ${\hat B}$ contain, in addition to the weak coupling constants, 
the particular isospin
structure corresponding to the exchanged meson. 

In the case of vector meson exchange the weak potential takes the form

\begin{eqnarray}
{V_v^{w} }({\vec q}\,)  &=&
G_F m_\pi^2
 \left( F_1 {\hat \alpha} - \frac{({\hat \alpha} + {\hat \beta} )
 ( F_1 + F_2 )} {4M_N \overline{M}}
%(\mbox{\boldmath $\sigma$}_1 \times {\bf q})
(\mbox{$\vec \sigma$}_1 \times {\vec q}\,)
%(\mbox{\boldmath $\sigma$}_2 \times {\bf q}) \right. \nonumber \\
(\mbox{$\vec \sigma$}_2 \times {\vec q}\,) \right. \nonumber \\
& & \phantom { G_F m_\pi^2 A }
\left. +i \frac{{\hat \varepsilon} ( F_1 + F_2 )} {2M_N}
%(\mbox{\boldmath $\sigma$}_1 \times
(\mbox{$\vec \sigma$}_1 \times
%\mbox{\boldmath $\sigma$}_2 ) {\bf q}\right)
\mbox{$\vec \sigma$}_2 ) {\vec q}\,\right)
\frac{1}{{\vec q\,}^2 + \mu^2} \, \ ,
\label{eq:vpot}
\end{eqnarray}
%with $\mu = m_v$ the mass of the meson, 
%$F_1 = g^{\rm {\scriptscriptstyle V}}_{\rm
% {\scriptscriptstyle NNV}}$,
%$F_2 = g^{\rm {\scriptscriptstyle T}}_{\rm {\scriptscriptstyle NNV}}$
with $F_1 = g^{\rm {\scriptscriptstyle V}}_{\rm
 {\scriptscriptstyle NNM}}$,
$F_2 = g^{\rm {\scriptscriptstyle T}}_{\rm {\scriptscriptstyle NNM}}$
the strong coupling constants and ${\hat \alpha}$, ${\hat
\beta}$ and ${\hat \varepsilon}$ the weak coupling constants that also contain
the appropriate isospin operator of the particular meson.

Performing a Fourier transform of Eqs. (\ref{eq:pspot}) and (\ref{eq:vpot}), and 
using the relation
%$({\mbox{\boldmath $\sigma$}}_1 \times {\bf q}) 
$({\mbox{$\vec \sigma$}}_1 \times {\vec q}) 
%({\mbox{\boldmath $\sigma$}}_2
({\mbox{$\vec \sigma$}}_2
\times {\vec q}\,) =
%({\mbox{\boldmath $\sigma$}}_1 {\mbox{\boldmath $\sigma$}}_2)
({\mbox{$\vec \sigma$}}_1 {\mbox{$\vec \sigma$}}_2)
%\: {\bf q}^2 - ({\mbox{\boldmath $\sigma$}}_1 {\bf q})
\: {\vec q\,}^2 - ({\mbox{$\vec \sigma$}}_1 {\vec q}\,)
% ({\mbox{\boldmath $\sigma$}}_2 {\bf q})$
 ({\mbox{$\vec \sigma$}}_2 {\vec q}\,)$
in (\ref{eq:vpot}), 
one obtains the weak
transition potential in coordinate space, which can be cast into the form

\begin{eqnarray}
V^{w} ({\vec r}\,) &=& \sum_{i} \sum_\alpha V_\alpha^{(i)}
({\vec r}\,) = \sum_i \sum_{\alpha}
V_\alpha^{(i)} (r) V_{\alpha}^{(i)}({\hat r}) \hat{I}_\alpha^{(i)} \nonumber
\\
&=& \sum_{i} \left[ V_C^{(i)}(r) \hat{I}^{(i)}_C + V_{SS}^{(i)}(r)
%\mbox{\boldmath $\sigma$}_1
%\mbox{\boldmath $\sigma$}_2 \hat{I}^{(i)}_{SS}
\mbox{$\vec \sigma$}_1
\mbox{$\vec \sigma$}_2 \hat{I}^{(i)}_{SS}
+ V_T^{(i)}(r)
S_{12}(\hat{r}) \hat{I}^{(i)}_T + \right. \nonumber \\
%& & \left. + \left( n^i \mbox{\boldmath $\sigma$}_2
%\cdot \hat{r}
& & \left. + \left( n^i \mbox{$\vec \sigma$}_2
\cdot \hat{r}
%+ (1-n^i) \left[\mbox{\boldmath $\sigma$}_1 \times
%\mbox{\boldmath $\sigma$}_2 \right] \cdot \hat{\bf r} \right)
+ (1-n^i) \left[\mbox{$\vec \sigma$}_1 \times
\mbox{$\vec \sigma$}_2 \right] \cdot \hat{r} \right)
V_{PV}^{(i)}(r) \hat{I}^{(i)}_{PV} \right] \ ,
\label{eq:genpot}
\end{eqnarray}
where the index $i$ runs over the different mesons exchanged ($i=1,\dots,
6$ represents $\pi,\eta,$K,$\rho,\omega,$K$^*$)
and
$\alpha$ over the different
spin operators denoted by $C$
(central spin independent), $SS$ (central spin dependent), $T$
(tensor) and $PV$ (parity violating). In the above expression,
$n^i = 1 (0)$ refers to the pseudoscalar (vector) mesons.
In the case of isovector mesons ($\pi$, $\rho$) the
isospin factor is
%$\mbox{\boldmath $\tau$}_1
%\mbox{\boldmath $\tau$}_2$ and for
$\mbox{$\vec \tau$}_1
\mbox{$\vec \tau$}_2$ and for
isoscalar mesons
($\eta$,$\omega$)
this factor is just $\hat{1}$ for all spin structure pieces of the potential.
In the case of isodoublet mesons (K,
K$^*$) there
are contributions proportional to $\hat{1}$ and to 
%$\mbox{\boldmath $\tau$}_1 \mbox{\boldmath $\tau$}_2$ 
$\mbox{$\vec \tau$}_1 \mbox{$\vec \tau$}_2$ 
that depend on the
coupling constants and, therefore, on the
spin structure piece of the potential denoted by $\alpha$.
For K-exchange we have
\begin{eqnarray}
{\hat I}_{C}^{(3)}&=& 0 \nonumber \\
{\hat I}_{SS}^{(3)}&=& {\hat I}_{T}^{(3)} =
\frac{
C^{\rm \scriptscriptstyle{P C}}_{\rm\scriptscriptstyle{K}}} {2} +
D^{\rm \scriptscriptstyle{P C}}_{\rm\scriptscriptstyle{K}} + \frac{
C^{\rm \scriptscriptstyle{P C}}_{\rm\scriptscriptstyle{K}}} {2}
%\mbox{\boldmath $\tau$}_1
%\mbox{\boldmath $\tau$}_2 \nonumber \\
\mbox{$\vec \tau$}_1
\mbox{$\vec \tau$}_2 \nonumber \\
{\hat I}_{PV}^{(3)} &=&
\frac{
C^{\rm \scriptscriptstyle{P V}}_{\rm\scriptscriptstyle{K}}} {2} +
D^{\rm \scriptscriptstyle{P V}}_{\rm\scriptscriptstyle{K}} + \frac{
C^{\rm \scriptscriptstyle{P V}}_{\rm\scriptscriptstyle{K}}} {2}
%\mbox{\boldmath $\tau$}_1
%\mbox{\boldmath $\tau$}_2 \, \ ,
\mbox{$\vec \tau$}_1
\mbox{$\vec \tau$}_2 \, \ ,
\end{eqnarray}
and for K$^*$-exchange
\begin{eqnarray}
{\hat I}_{C}^{(6)} &=&  \frac{
C^{\rm \scriptscriptstyle{PC,V}}_{\rm\scriptscriptstyle{K^*}}} {2} +
D^{\rm \scriptscriptstyle{PC,V}}_{\rm\scriptscriptstyle{K^*}} + \frac{
C^{\rm \scriptscriptstyle{PC,V}}_{\rm\scriptscriptstyle{K^*}}} {2}
%\mbox{\boldmath $\tau$}_1
%\mbox{\boldmath $\tau$}_2 \nonumber \\
\mbox{$\vec \tau$}_1
\mbox{$\vec \tau$}_2 \nonumber \\
{\hat I}_{SS}^{(6)} &=& {\hat I}_{T}^{(6)} =
\frac { \left(
C^{\rm \scriptscriptstyle{PC,V}}_{\rm\scriptscriptstyle{K^*}} +
C^{\rm \scriptscriptstyle{PC,T}}_{\rm\scriptscriptstyle{K^*}} \right)} {2}
+ \left( D^{\rm \scriptscriptstyle{PC,V}}_{\rm\scriptscriptstyle{K^*}} +
D^{\rm \scriptscriptstyle{PC,T}}_{\rm\scriptscriptstyle{K^*}} \right) +
\frac{ \left( C^{\rm \scriptscriptstyle{PC,V}}_{\rm\scriptscriptstyle{K^*}} +
C^{\rm \scriptscriptstyle{PC,T}}_{\rm\scriptscriptstyle{K^*}} \right) }{2}
%\mbox{\boldmath $\tau$}_1
%\mbox{\boldmath $\tau$}_2 \nonumber \\
\mbox{$\vec \tau$}_1
\mbox{$\vec \tau$}_2 \nonumber \\
{\hat I}_{PV}^{(6)} &=&
\frac{
C^{\rm \scriptscriptstyle{PV}}_{\rm\scriptscriptstyle{K^*}}} {2} +
D^{\rm \scriptscriptstyle{PV}}_{\rm\scriptscriptstyle{K^*}} + \frac {
C^{\rm \scriptscriptstyle{PV}}_{\rm\scriptscriptstyle{K^*}}} {2}
%\mbox{\boldmath $\tau$}_1
%\mbox{\boldmath $\tau$}_2  \ .
\mbox{$\vec \tau$}_1
\mbox{$\vec \tau$}_2  \ .
\end{eqnarray}

The different pieces $V_{\alpha}^{(i)}(r)$, with $\alpha=C,SS,T,PV$,
are given by
\begin{eqnarray}
V_{C}^{(i)} (r) &=&  K^{(i)}_{C}
\frac {{\rm e}^{- \mu_i r}} {4 \pi r} \equiv K^{(i)}_{C} \:
V_{C} (r,\mu_i)
\label{eq:cpot}
 \\
V_{SS}^{(i)}(r) &=& K^{(i)}_{SS}  \frac{1}{3}
\: \left[ {\mu_i}^2  \: \frac {{\rm e}^{- \mu_i r}} {4 \pi r}
- \delta ({\vec r}\,) \right] \equiv K^{(i)}_{SS} V_{SS} (r,\mu_i)
\label{eq:sspot} \\
V^{(i)}_{T} (r) &=& K^{(i)}_{T} \: \frac {1}{3}
\:  \mu_i^2  \: \frac {{\rm e}^{- \mu_i r}} {4 \pi r} \:
\left( 1 + \frac{3} {{\mu_i} r} + \frac{3} {({\mu_i
r})^2} \right)
 \equiv K^{(i)}_{T} V_{T} (r,\mu_i)
\label{eq:tpot}
 \\
V_{PV}^{(i)}(r) &=& K^{(i)}_{PV} \: \mu_i \:
\frac {{\rm e}^{- {\mu_i} r}} {4 \pi r}
\left( 1 + \frac{1} {\mu_i r} \right)
 \equiv  K^{(i)}_{PV} V_{PV}(r,\mu_i)  \ .
\label{eq:pvpot}
\end{eqnarray}
where $\mu_i$ denotes the mass of the
different mesons.
%It is these expressions that are inserted in Eq.
%(\ref{eq:cs3}) to compute the two-body transition matrix
%elements numerically.
The expressions for $K^{(i)}_\alpha$, which contain factors and
coupling constants, are given in Table
\ref{tab:ctspot}. The explicit values of the coupling constants can be found
in Table \ref{wkp} of Ref. \cite{PRB97}. 

A monopole form factor $F_{i}({\vec q\,}^2)=
(\Lambda_i^2-\mu_i^2)/(\Lambda_i^2+{\vec q\,} ^2)$
is used at each vertex, where the value of the cut-off,
$\Lambda_i$, 
depends on the meson. 
These values are the same ones as those of the strong J\"ulich 
$YN$ interaction\cite{HHS89}, 
since the Nijmegen model distinguishes form
factors only in terms of the transition channel. The resulting expressions 
for the regularized potentials, which were given already in Ref. \cite{PRB97},
will be repeated here in order to correct for some misprints. The effect
of form factors is included by making the following replacements in 
Eqs. (\ref{eq:cpot}) to (\ref{eq:pvpot})
\begin{eqnarray}
V_{\rm C} (r; \mu_i) &\to& V_{\rm C} (r; \mu_i) - V_{\rm C}
(r;\Lambda_i) -
\frac{ {\Lambda_i}^2 - {\mu_i}^2}{2 \, \Lambda_i} \frac{{\rm e}^{-
\Lambda_i r}}{4 \pi}
\label{eq:cregpot} \\
V_{\rm SS} (r; \mu_i) &\to& V_{\rm SS} (r; \mu_i) - V_{\rm SS}
(r;\Lambda_i) -
\Lambda_i \frac{ {\Lambda_i}^2 - {\mu_i}^2}{6} \frac{{\rm e}^{-
\Lambda_i r}}{4 \pi}
\left( 1 - \frac{2}{\Lambda_i r} \right)
\label{eq:ssregpot} \\
V_{\rm T} (r; \mu_i) &\to& V_{\rm T} (r; \mu_i) - V_{\rm T}
(r; \Lambda_i) -
\Lambda_i \frac{ {\Lambda_i}^2 - {\mu_i}^2}{6} \frac{{\rm e}^{-
\Lambda_i r}}{4 \pi}
\left( 1 + \frac{1}{\Lambda_i r} \right)
\label{eq:tregpot} \\
V_{\rm PV} (r; \mu_i) &\to& V_{\rm PV} (r; \mu_i) - V_{\rm PV}
(r; \Lambda_i) -
\frac{ {\Lambda_i}^2 - {\mu_i}^2}{2} \frac{{\rm e}^{- \Lambda_i r}}
{4\pi} \ .
\label{eq:pvregpot}
\end{eqnarray}

\section{Results}
\label{res}

\begin{figure}
   \caption{Total cross sections for the reaction $p n \to p \Lambda$ as a
function of the proton lab momentum using the weak one-pion-exchange
potential. Dashed line: calculation omitting form factors and strong
correlations; thin solid line: including form factors and $NN$ initial
correlations; thick solid line: including form factors and both $NN$
and $\LN$ distortions. }
\vspace{-4cm}
       \setlength{\unitlength}{1mm}
       \begin{picture}(100,180)
       \put(5,-40){\epsfxsize=13cm \epsfbox{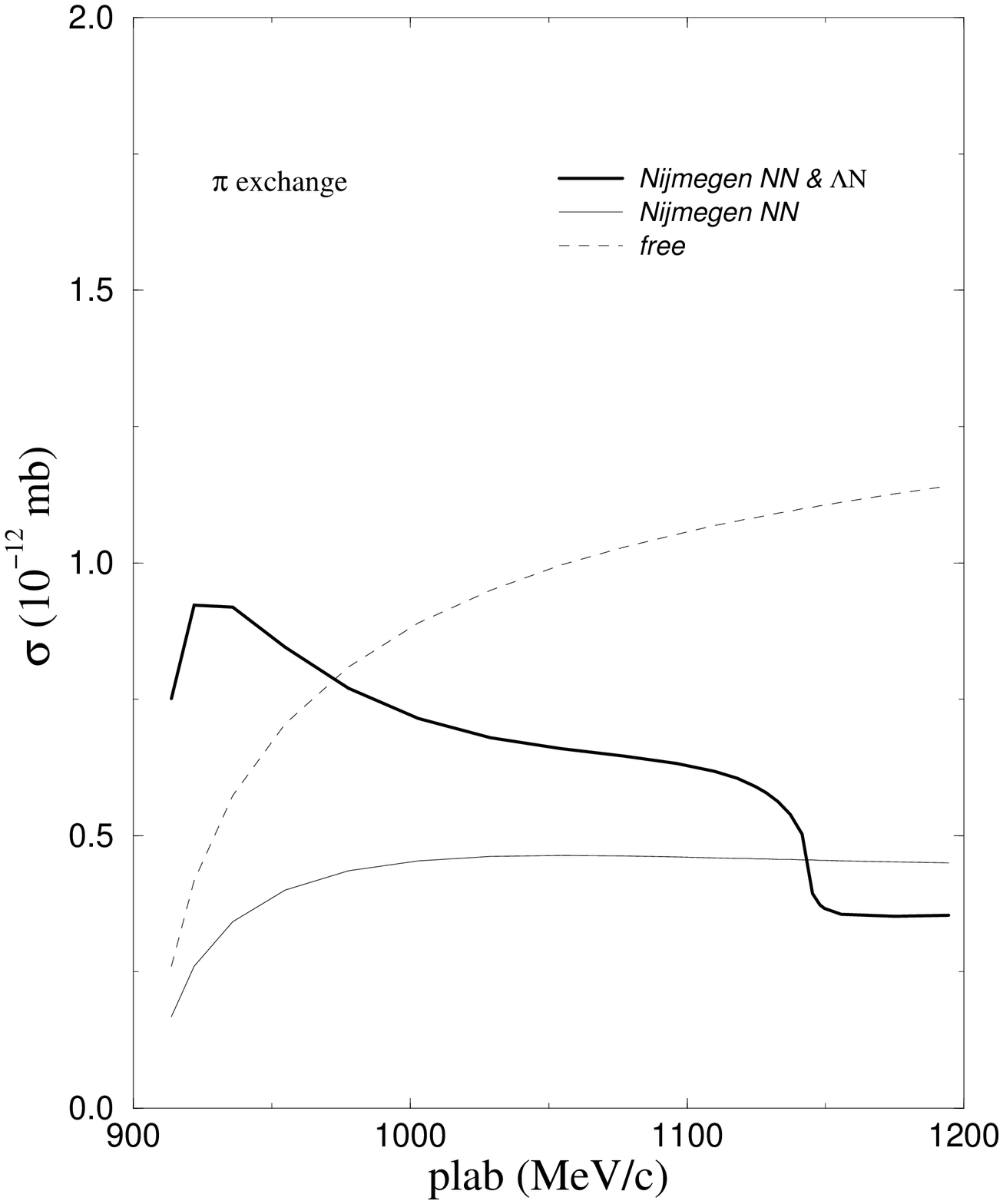}}
       \end{picture}
   \label{fig:fig1}
\end{figure}

The total cross section for the $p n \to p \Lambda$ reaction including only
one-pion-exchange in the weak transition potential is shown in
Fig.~\ref{fig:fig1} as a function of the proton lab momentum, starting at the
$p \Lambda$ threshold. As expected for a weak process,
the cross section is of the order of 10$^{-12}$mb. 
The dashed line corresponds to the ``free" calculation where
form factors are omitted at the vertices and
the incoming $pn$ and outgoing $p\Lambda$ states propagate as plane waves.
The thin solid line includes form factors and 
strong initial $pn$ distorted waves obtained from the Nijmegen
93 $NN$
interaction \cite{nijmnn}. The effect of the $NN$ interaction is to reduce 
the cross section by a factor of about 2. 
The reduction is a result of the fact that the $pn$ pair moves with large
relative momentum (from $415$ MeV/c to $525$ MeV/c) in the range of proton
lab momenta explored and feels the effect of the strong
repulsive short range part of the $NN$ interaction. As a consequence,
the $pn$ wave function is much reduced at small distances where the weak 
transition potential contributes most to the cross section and hence 
the cross section becomes smaller. 
The thick solid line shows the cross section when the $p \Lambda$
distortions, obtained from the $YN$ Nijmegen Soft Core potential
\cite{nijmln}, are also included. When compared
to the thin line, one observes a substantial enhancement of the
cross section
in the low momentum region which turns into a moderate reduction
at high lab momentum values. This is due to the fact that, close to
threshold,
the $p \Lambda$ pair moves with  small relative momentum, feeling
 the attractive component of the $YN$ potential. As a
result the wave function gets pushed in to smaller distances 
and the cross section is enhanced.
As the proton lab momentum increases, so does the relative $p
\Lambda$ momentum and, eventually, the repulsive core of the $YN$
interaction becomes responsible for reducing the wave function
at short distances, giving rise to smaller cross sections.
The opening of the $\Sigma N$ channel, coupled to the $\Lambda N$ one
through the strong interaction,
shows up as a step in the cross section at a proton lab momentum 
around $1140$ MeV/c.

\begin{figure}
   \caption{Total cross sections for the reaction $p n \to p \Lambda$ as a
function of the proton lab momentum. The strong distortions are
generated with the $NN$ Nijmegen 93 and $YN$ Nijmegen Soft Core models (a)
or by the $NN$ Bonn B and $YN$ J\"ulich A (b).
Dashed line: $\pi$-exchange only; thin solid line: $\pi + \rho$;
thick solid line: full set of mesons.}
\vspace{-9cm}
       \setlength{\unitlength}{1mm}
       \begin{picture}(100,180)
       \put(3,-40){\epsfxsize=12cm \epsfbox{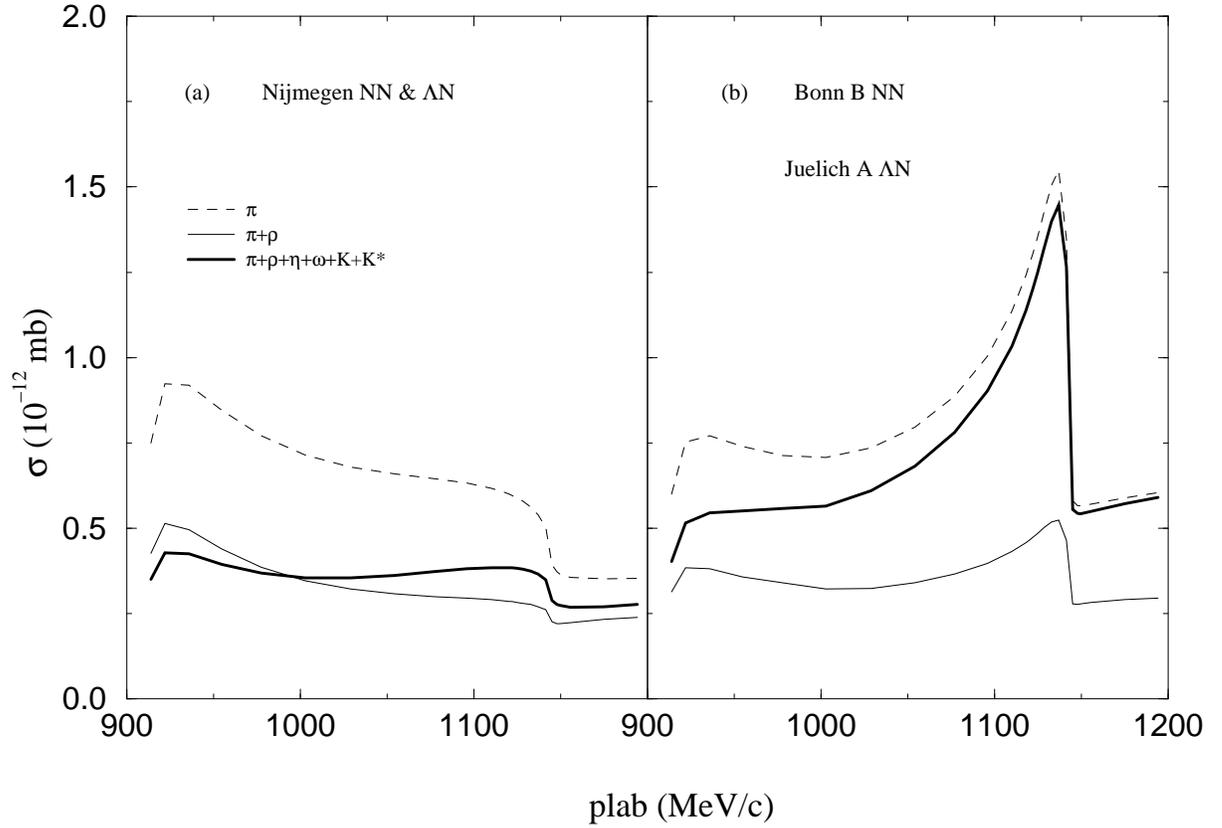}}
       \end{picture}
   \label{fig:fig2}
\end{figure}

The effect of including the different mesons in the weak transition
potential is shown in Fig.~\ref{fig:fig2}. Fig.~\ref{fig:fig2}a shows the cross sections
calculated using the Nijmegen potentials for getting the $NN$ and $\LN$
distorted waves and Fig.~\ref{fig:fig2}b shows those using the Bonn B $NN$ and J\"ulich A 
$\LN$ potentials. In Fig.~\ref{fig:fig2}a we see that adding the $\rho$
meson to one-pion-exchange considerably reduces the cross section
in the kinematic region studied here. Including the
remaining mesons does not change the cross section much, which ends up
being about a factor 2 smaller than that for one-pion only. This effect
appears to be surprisingly different from
the moderate reduction of 15\%
found for the decay rate of hypernuclei when the effect of adding heavier
mesons to the one-pion-exchange mechanism was studied \cite{PRB97}.
However, we point out that the $\LN$ correlations
in the weak hypernuclear decay case were obtained from a G-matrix
calculation which takes into account the Pauli principle acting on the
intermediate nucleon. This blocks the low relative momentum
transitions between the initial $\LN$  and intermediate $YN$
states and, therefore, the effect of the attractive part of the
interaction on the correlated $\LN$ wave function is much 
reduced compared to those of the distorted $p
\Lambda$ wave function in free space. Thus, the $\LN$
wave function in the hypernucleus is strongly suppressed at 
small distances, making the contribution of the heavier mesons less
important. 

As can be seen in Fig.~\ref{fig:fig2}b, the use of the Bonn B $NN$ and J\"ulich A $\LN$
potentials give rise to different effects
 in the cross section as compared to those calculated
using the Nijmegen ones. Although not
shown in the figures, we have checked that, in the absence of $p
\Lambda$ distortions,
the cross section calculated with the distorted $p n$ waves,
obtained from either the Nijmegen 93 or the Bonn B potentials, 
yield nearly identical results.
Therefore, the differences between the results in
Figs.~\ref{fig:fig2}a and~\ref{fig:fig2}b come almost completely from the
different models used to distort the $p \Lambda$ final state.
Already at the level of only pion-exchange (dashed line in Fig.~\ref{fig:fig2}b),
the J\"ulich A model shows a clear enhancement close
to the $p \Sigma$ threshold. Adding the $\rho$ meson furthermore yields
 a strong reduction of the cross section.
On the other hand, the addition of the remaining mesons produces a
substantial
enhancement which gives rise to a final cross section not very different
from the pion-only result.
We note here that the pion-only cross section shown in Fig.~\ref{fig:fig2}
is close to the results obtained in Ref.
\cite{HHKST95} with one-pion-exchange, but we do not find the tremendous 
enhancement of
a factor of 3 in the cross section when the 
$\rho$ meson is included.
This is most likely due to the different models for the weak
$\rho N \Lambda$ vertices used in the
 $\rho$-exchange mechanism. Their weak coupling
constants are larger than 
the ones used here by a factor of more  than 2 for the parity
conserving ones ($\alpha_\rho$
and $\beta_\rho$) and a factor 3.5 for the parity violating one
($\varepsilon_\rho$). 

\begin{figure}
   \caption{Different partial wave contributions to the cross section
for the $p n \to p \Lambda$ reaction. Only the pion is considered in
the weak transition potential. The $NN$ and $\LN$ distorted
waves are generated with the Nijmegen models (a) and the Bonn B and 
J\"ulich A models (b) respectively.}
\vspace{-9cm}
       \setlength{\unitlength}{1mm}
       \begin{picture}(100,180)
       \put(3,-40){\epsfxsize=12cm \epsfbox{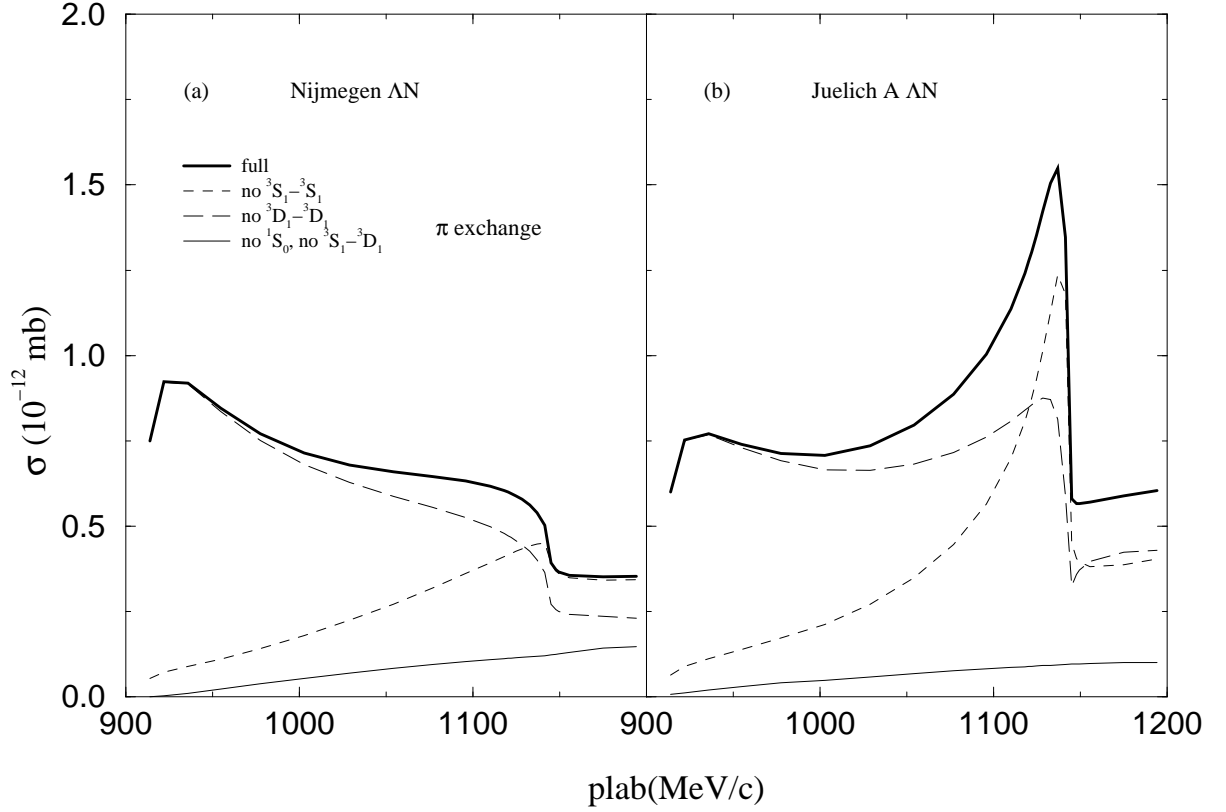}}
       \end{picture}
   \label{fig:fig3}
\end{figure}

Let us now explore the origin of the peak in the cross section around the 
$\Sigma$ threshold.
Nuclear matter calculations already
showed some time ago \cite{YRHNM92,YMHIN94,HPRM96} that, while the
$\Lambda$ binding energy in nuclear matter turned out to be about
$-25$ to $-30$ MeV for the Nijmegen and J\"ulich interactions, the
distribution of this strength between 
the various partial waves was very different. The
most important contribution for the J\"ulich interactions came from
the coupled $^3S_1-^3D_1$ channel, the $^1S_0$ contribution being
negligible. However, the Nijmegen interactions obtained almost equal
contributions from both the $J=0$ and $J=1$ channels.
In Fig.~\ref{fig:fig3} we analyze the contribution to the $p n \to p
\Lambda$ cross section from  
$^3S_1-^3D_1$ partial waves obtained from
the Nijmegen (Fig.~\ref{fig:fig3}a) and J\"ulich A (Fig.~\ref{fig:fig3}b) 
$\Lambda N$ interactions. The
thick solid lines (``{\em full} '') are the calculated cross sections with
one-pion-exchange for the weak transition $p n \rightarrow p\Lambda$ and
including distorted waves for the $pn$ and $p\Lambda$ systems. The  
short dashed lines represent cross sections obtained by removing
the $^3S_1 \to ^3S_1$ component of the distorted $p \Lambda$ 
wave function and the 
long-dashed lines those obtained when the $^3D_1 \to ^3D_1$ component is
removed. 
The thin solid line shows the small cross sections obtained when all the
$^3S_1-^3D_1$ coupled channel is omitted, thus reflecting the fact
that these are the most important components
for the $p n \rightarrow
p \Lambda$ reaction, as already noted in Ref. \cite{HHKST95}. 
The stronger $^3S_1-^3D_1$ channel of the J\"ulich interaction, relative
to the Nijmegen one, is the reason for the much more enhanced cross section
at the $N \Sigma$ threshold.
Suppressing the $^1S_0$ partial waves barely affects the cross section
and cannot be seen in Figs.~\ref{fig:fig3}a and~\ref{fig:fig3}b. 
In spite of the fact that the
Nijmegen
$YN$ interaction has a sizable $^1S_0$ component, its contribution to
the cross section gets eliminated due to the weak one-pion-exchange
transition potential which has negligible
strength for the $^1S_0 \to ^1S_0$ transition.
 
\begin{figure}
   \caption{Differential cross sections in the center of mass for the 
reaction $p n \to p
\Lambda$ at $p_{\rm lab}=1137$ MeV/c.
Dashed line: $\pi$-exchange only; thin solid line: $\pi + \rho$;
thick solid line: full set of mesons. The $NN$ and $\LN$ wave functions
are generated using Bonn B and J\"ulich A models respectively.}
\vspace{-4cm}
       \setlength{\unitlength}{1mm}
       \begin{picture}(100,180)
       \put(5,-40){\epsfxsize=13cm \epsfbox{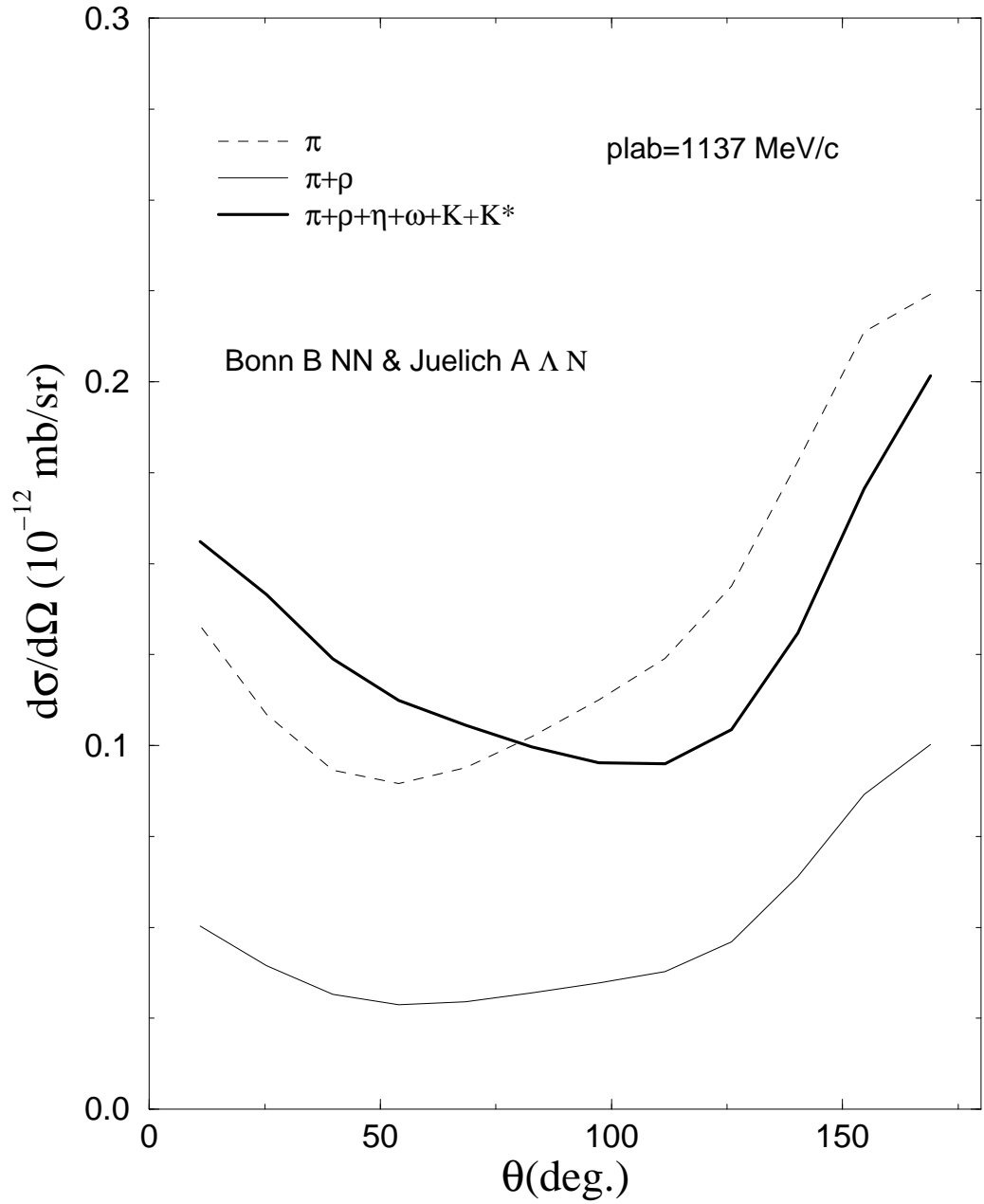}}
       \end{picture}
\label{fig:fig4}
\end{figure}

In Fig.~\ref{fig:fig4} we show the differential cross
sections for the proton lab momentum of $1137$ MeV/c which
corresponds to the position of the peak in Fig.~\ref{fig:fig2},
where the
strong $NN$ and $\LN$ correlations have been generated with the
Bonn B and J\"ulich A interactions, respectively.
The dashed line is the result for the pion-only calculation, the
thin solid line shows the effect of including the $\rho$ meson
and the thick solid line incorporates the full set of six mesons.
The effect of adding the rho to the pion is a reduction of a factor 
slightly more than 2 while adding all mesons brings the results closer 
to the one-pion
ones, as observed earlier for the total cross section.

\begin{figure}
   \caption{Total asymmetry $A$ for the reaction $p n \to
p \Lambda$
as a function of the proton lab momentum using the weak one-pion-exchange
potential. Dashed line: calculation omitting form factors and strong
correlations; thin solid line: including form factors and $NN$ distortions; 
thick solid line: including form factors and both $NN$
and $\LN$ distortions. }
\vspace{-4cm}
       \setlength{\unitlength}{1mm}
       \begin{picture}(100,180)
       \put(5,-40){\epsfxsize=13cm \epsfbox{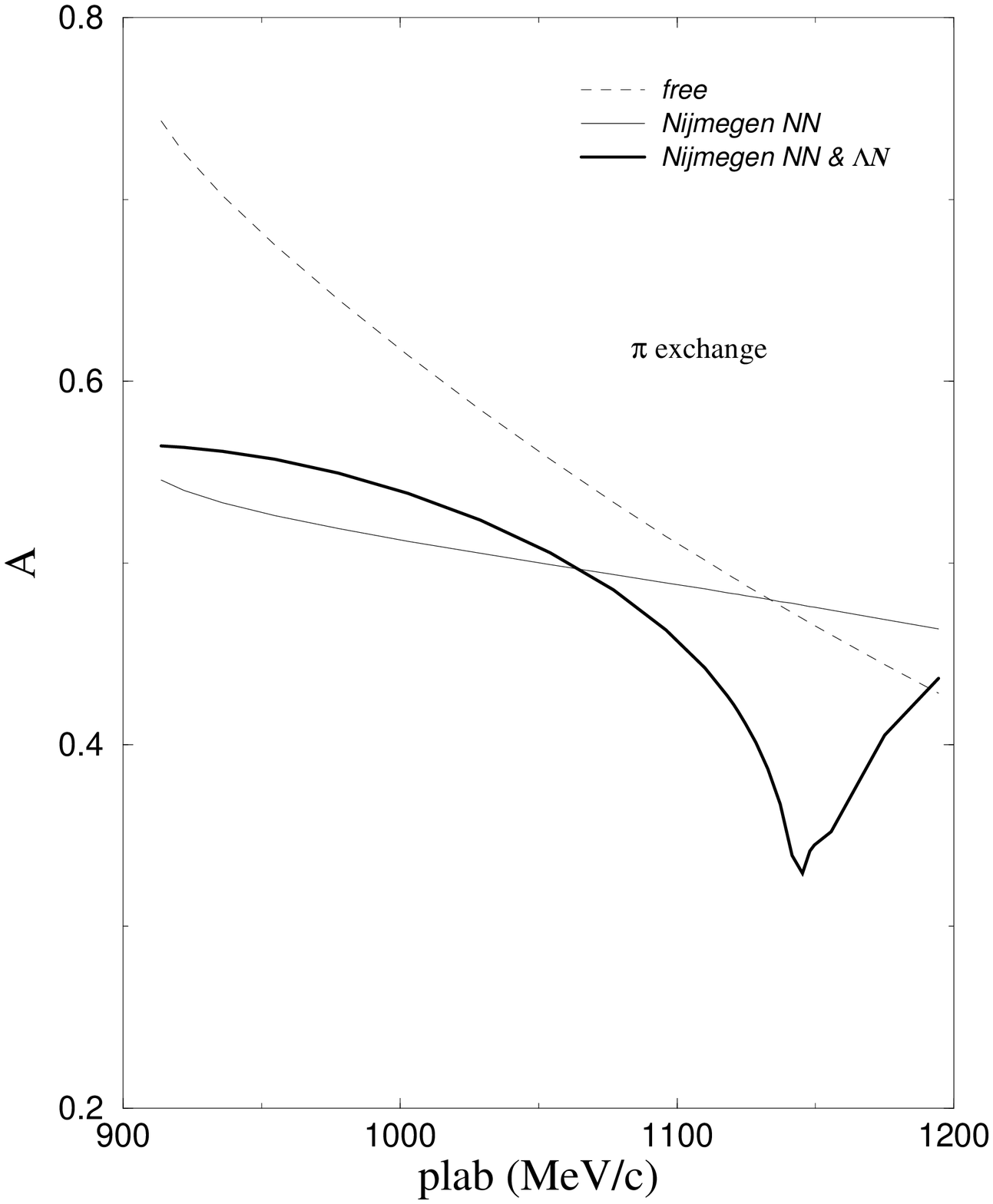}}
       \end{picture}
   \label{fig:fig5}
\end{figure}

A measure of the amount of parity violation in the weak $\Lambda$
production is given by the asymmetry $A$ defined as
\begin{equation}
A=\frac{\sigma_+ - \sigma_-}{\sigma_+ + \sigma_-} \ ,
\end{equation}
where $\sigma_+$ ($\sigma_-$) is the cross section for positive
(negative) helicity of the incoming proton.
In Fig.~\ref{fig:fig5} we consider the one-pion-exchange mechanism in
the weak $p n \to p \Lambda$ transition and illustrate the
effect of the strong distortions on the asymmtery A using the
Nijmegen models for the $NN$ and $\LN$ interactions.
The dashed line shows the asymmetry when neither the  
form factors at the vertices nor the strong distortions are included. The
inclusion of the $NN$ interactions, as shown by the thin solid line, 
decrease the value of the asymmetry up
to a beam momentum of 1130 MeV/c
beyond which it increases compared to the dashed line. 
The addition of the $\LN$ final state
interaction gives rise to the particular cusp structure due to the
opening of the $\Sigma N$ channel around 1140 MeV/c.

\begin{figure}
   \caption{Total asymmetry for the reaction $p n \to p
\Lambda$ as a
function of the proton lab momentum. The $NN$ and $\LN$ distortions are
generated with the Bonn B and J\"ulich A potentials
respectively.
Dashed line: $\pi$-exchange only; thin solid line: $\pi + \rho$;
thick solid line: full set of mesons.}
\vspace{-4cm}
       \setlength{\unitlength}{1mm}
       \begin{picture}(100,180)
       \put(5,-40){\epsfxsize=13cm \epsfbox{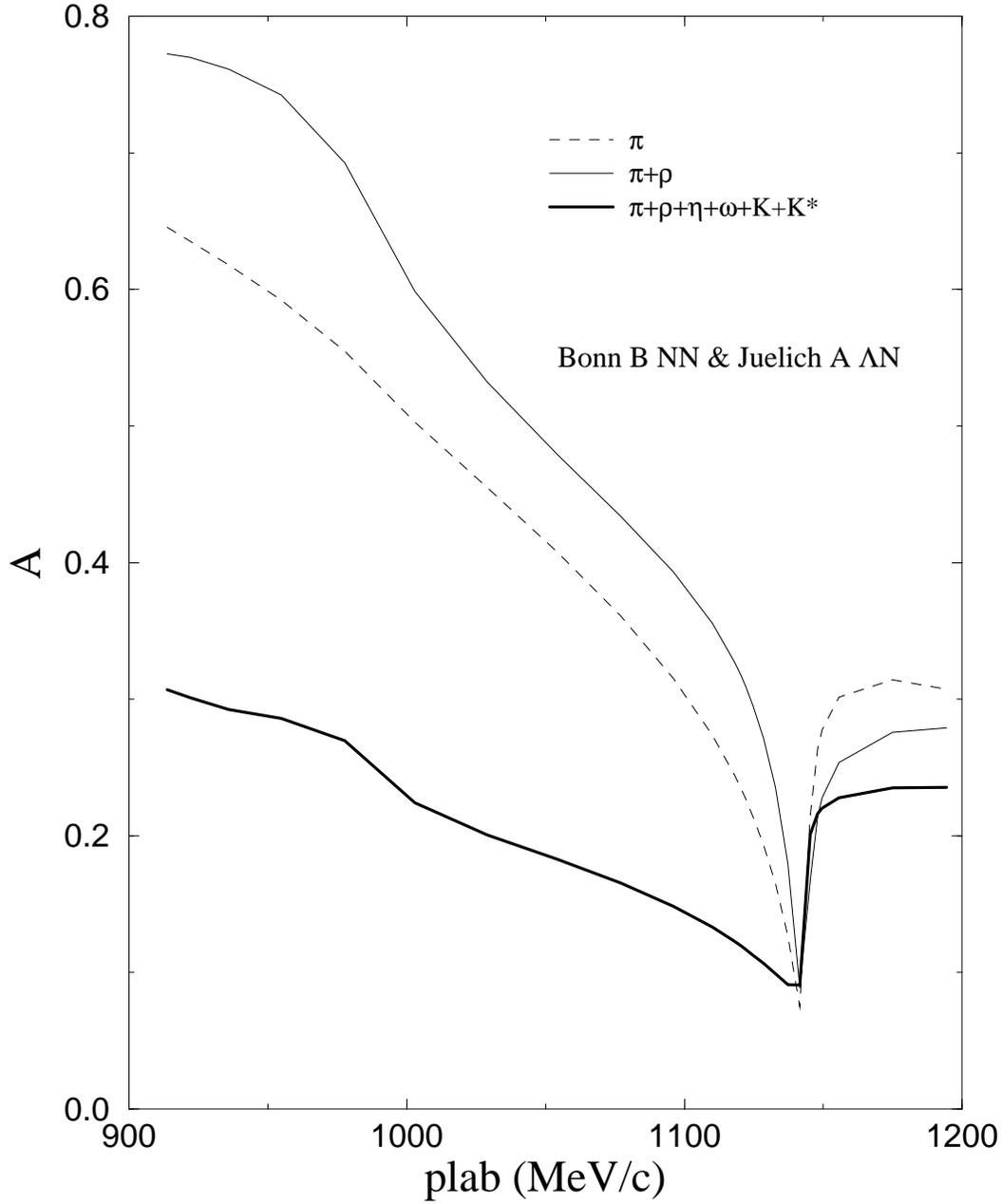}}
       \end{picture}
   \label{fig:fig6}
\end{figure}

The effect of the different meson exchanges in the weak transition
potential is shown in Fig.~\ref{fig:fig6} where the $NN$ and $\LN$ distorted
waves have been generated with the Bonn B and J\"ulich A models, 
respectively. 
We observe
that while the $\rho$ meson increases the asymmetry with respect to
the pion-only result, the inclusion of all the six mesons produces an
asymmetry which is considerably smaller. Similar qualitative features
are found when the Nijmegen potential models are used in constructing
the distorted $NN$ and $\LN$ waves.

\begin{figure}
   \caption{Total cross section (a) and asymmetry (b) for the
reaction $p n \to p \Lambda$ as a
function of the proton lab momentum using the full meson-exchange
weak transition potential with various strong potential models.
Dashed line: $NN$ Nijmegen 93 and $YN$ Nijmegen soft-core;
thin solid line: $NN$ Bonn B and $YN$ J\"ulich B;
thick solid line: $NN$ Bonn B and $YN$ J\"ulich A.}
\vspace{-9cm}
       \setlength{\unitlength}{1mm}
       \begin{picture}(100,180)
       \put(-6,-40){\epsfxsize=12cm \epsfbox{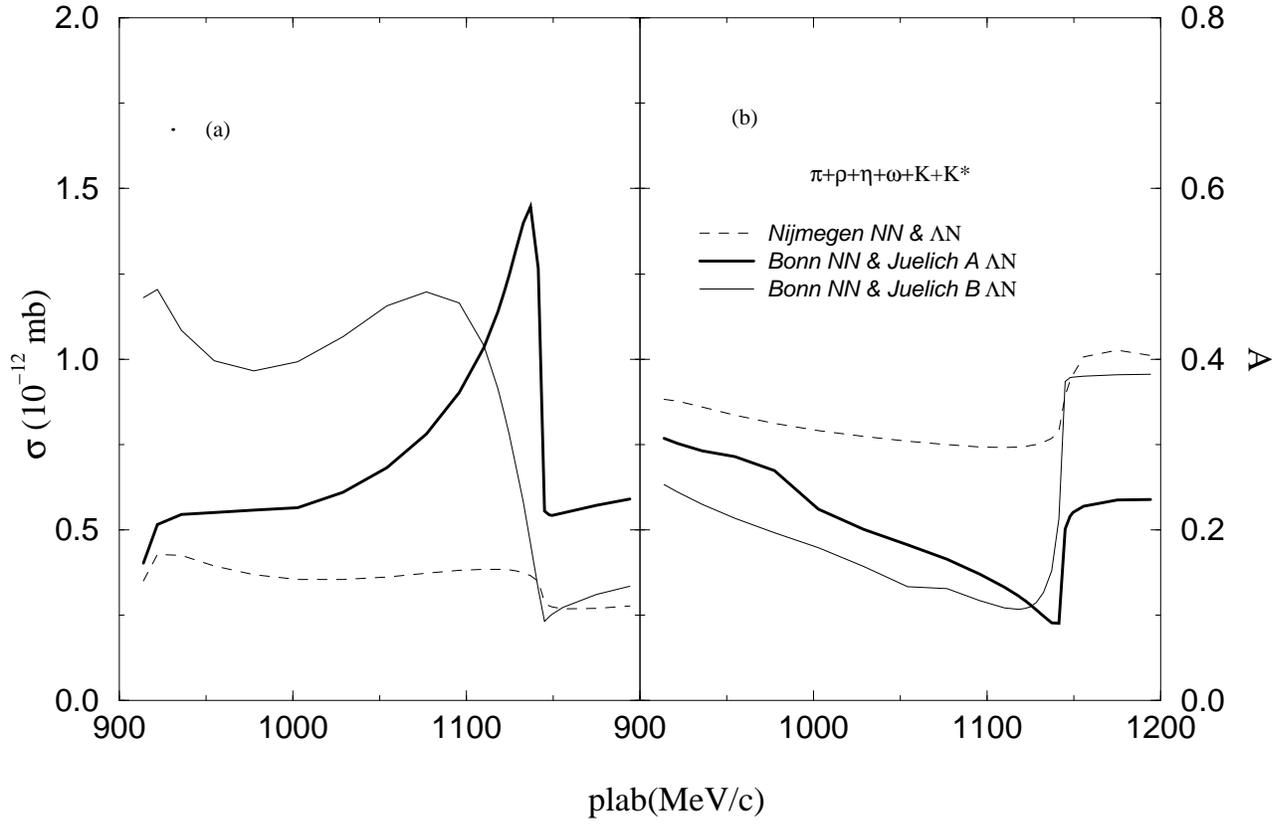}}
       \end{picture}
   \label{fig:fig7}
\end{figure}

Finally, we present in Figs.~\ref{fig:fig7}a and~\ref{fig:fig7}b the 
$p n \to p \Lambda$ cross
sections 
and asymmetries, respectively,
obtained with the full meson-exchange weak transition potential for
three different strong interaction models. 
The dashed lines use the $NN$ Nijmegen 93 and $YN$
Nijmegen Soft Core models, the thin solid lines use the $NN$ Bonn B and
$YN$ J\"ulich B models and the thick solid lines use the $NN$ Bonn
B and the $YN$ J\"ulich A models. As mentioned before, the differences
between the curves are mainly due to the differences in the $YN$ potentials
employed. 

It is clear that these strong $YN$ potentials are not sufficiently
constrained by the small amount of total cross section
data on $YN$ scattering. Hence the different $YN$ models which produce
the total cross sections for $YN$ scattering equally well, 
give rise to very different
predictions when applied to other reactions that are sensitive to
the $YN$ interaction, like hypernuclear structure calculations, 
studies of nuclear matter with strangeness, or
the weak transition $p n \to p \Lambda$ studied here. 
More data on $YN$ scattering, especially on differential cross
sections and polarization observables, is highly desirable in order to
constraint
the $YN$ interactions sufficiently well so that the $p n \to p
\Lambda$ reaction can be used to learn about the weak four fermion
interaction. In particular, the two-step  
$p n \to N \Sigma \to p \Lambda$ transition, not considered in the
present work, could be studied after having a better knowledge of the
$YN$ interaction.

\section{Summary}
\label{sum}

We have studied the weak strangeness production reaction $p n \rightarrow 
p \Lambda$ in a distorted wave Born approximation formalism using the 
one-meson-exchange model for the weak transition consisting of six mesons
viz. the $\pi, \rho, \eta, \omega,$K and K$^*$. The distorted wave functions
are written in terms of partial wave expansions and are generated using the
different available potentials for nucleon-nucleon and hyperon-nucleon
interactions. 

The total cross sections are sensitive to the model ingredients of the
weak transition operator. Including the $\rho$ meson decreases the pion-only
cross sections by a factor of 2 or more. The effect of the 
remaining mesons depends on the strong potentials employed to distort
the $p \Lambda$ states, giving rise to cross sections that can be
either very close or a factor 2 smaller than the pion-only results. 

The kinematical region explored by the free $p n \rightarrow p \Lambda$
reaction is much larger than that by the inverse reaction, the
nonmesonic
decay $\Lambda N \rightarrow N N$ taking place inside hypernuclei. The 
heavier mesons contribute very differently and are more important
in the free reaction compared to
 the nonmesonic decay due to the different behavior of
the $\LN$ wave function inside a hypernucleus compared to that in free space.

The total cross section computed with the Nijmegen $\LN$ wave functions
show  a step-like behavior around 1140 MeV/c beam momentum, where the
strong $\Lambda N \rightarrow \Sigma N$ transition opens up. 
The J\"ulich $\LN$ results
show a dramatic peak in this region. However, the peak in our results is
not as
pronounced as the one found in earlier work\cite{HHKST95}.
The major contribution to these cross sections comes from the 
$\LN$ partial waves in the $^3S_1-^3D_1$ coupled channels.

We find the $p n \rightarrow p \Lambda$ reaction to be very sensitive to
the type of
model used for the strong hyperon-nucleon interaction. Hence, more data
on $YN$
scattering for observables other than the total cross sections are needed
 to constrain the $YN$ interaction models and use the 
$p n \rightarrow \Lambda p$ reaction to extract the weak four fermion
interaction.

The cross sections for the $p n \rightarrow p \Lambda$ reaction are of the
order of 10$^{-12}$ mb and are at the borderline of feasibility for the
existing experimental facilities.  

\section{Acknowledgements}
This work is partially supported by the DGICYT contract No. PB95-1249
(Spain), the Generalitat de Catalunya grant No. GRQ94-1022, the US-DOE grant 
No. DE-FG02-95-ER40907 and the NATO grant No. CRG960132. One of
the authors (N.G.K.) would like to thank the warm hospitality of the
nuclear physics group at the University of Barcelona. 

%\begin{table}               
%\bigskip
%\bigskip
%\caption{Dependence of the $\Lambda$ single-particle energy in $^{17}_\Lambda$O
%on the starting energy of the nuclear matter $G$-matrix. Our notation
%is $\omega= <B_N> + B_\Lambda(k=0)$, with $<B_N>=-50$ MeV.
%}
%\bigskip
%\bigskip
%\begin{tabular}{c| cc |cc}
%$k_F=1.36$ fm$^{-1}$ \phantom{caca}&
%\multicolumn{2}{c|}{Nijmegen}
%& \multicolumn{2}{c}{J\"ulich} \cr \hline
%$\omega$ \phantom{cac}& $HF$ & $HF + 2p1h$ \phantom{ca}& $HF$ & $HF + 2p1h$ \cr
%(MeV) \phantom{cac} & (MeV) & (MeV) \phantom{cac} & (MeV) & (MeV) \cr
%\hline
%$-100$ \phantom{cac}& $-3.83$ & $-7.43$ \phantom{cac}& $-9.25$ & $-11.85$ \cr
%$-80$  \phantom{cac}& $-4.76$ & $-7.39$ \phantom{cac}& $-10.15$ & $-11.83$ \cr
%$-50$  \phantom{cac}& $-5.59$ & $-7.36$ \phantom{cac}& $-11.73$ & $-11.84$ \cr
%\end{tabular}
%\label{tab:olambda}
%\end{table}

%      Figure 1, HF, 2p1h

\newpage

%\section{Appendix: The weak matrix elements}
\section{Appendix}
\label{wme}

The Lippmann-Schwinger equation allows us to obtain the 
scattered wave function for a pair of particles moving
under the influence of the strong interaction. The scattered states
will then be given by

\begin{equation}
|\Psi^{(+)}_{\sst NN } \rangle = |\Phi_{\sst NN } \rangle + 
\frac{1}{E_{\sst NN} - {H_0}_{\sst NN} + i \eta} T_{\sst NN}  
|\Phi_{\sst NN} \rangle  
\label{eq:wfnn1}
\end{equation}
for the incoming $NN$ states and by
\begin{equation}
\langle \Psi^{(-)}_{\sst \LN} | = \langle \Phi_{\sst \LN} | + 
\langle \Phi_{\sst \LN} |  T_{\sst \LN} \frac{1}{E_{\sst \LN} - 
{H_0}_{\sst \LN} + i \eta}
\label{eq:wfln1}
\end{equation}
for the outgoing $\LN$ states. The previous equations are written
in terms of the T-matrix which is obtained from 
\begin{equation}
T=V + V \frac{1}{E - H_0 + i \eta} T 
\end{equation}
and fulfills 
\begin{eqnarray}
& & V_{\sst NN} | \Psi^{(+)}_{\sst NN} \rangle=T_{\sst NN} |\Phi_{\sst NN} 
\rangle \\
\label{eq:tmat2}
& & V_{\sst \LN} | \Psi^{(-)}_{\sst \LN} \rangle=T^{\dagger}_{\sst \LN} |
\Phi_{\sst \LN} \rangle \, \ ,
\label{eq:tmat3}
\end{eqnarray}
$\mid \Phi_{\sst NN} \rangle$ and $\mid \Phi_{\sst \LN} \rangle$ being the 
corresponding
unperturbed states for the
$NN$ and $\LN$ systems, respectively. Note that Eq.~(\ref{eq:tmat3}) should 
have been written as 
\begin{equation}
V_{\sst \LN \to \LN} | \Psi^{(-)}_{\sst \LN \to \LN} \rangle +
V_{\sst \Sigma N \to \LN} | \Psi^{(-)}_{\sst \Sigma N \to \LN} \rangle 
=T^{\dagger}_{\sst \LN \to \LN} |
\Phi_{\sst \LN \to \LN} \rangle \, \ .
\label{eq:tmat32}
\end{equation}
However, since we disregard the $p n \to N \Sigma  \to p \Lambda$ transition, 
we don't need to consider the $\Sigma N \to \LN$ component of the wave function,
$| \Psi^{(-)}_{\sst \Sigma N \to \LN} \rangle $. Simplifying the notation
by writing $\LN \to \LN$ as simply $\LN$ Eq.~(\ref{eq:tmat32})
finally reduces to Eq.~(\ref{eq:tmat3}).

Projecting Eqns. (\ref{eq:wfnn1}) and (\ref{eq:wfln1}) into ${\vec r}$-space
and performing a partial-wave decomposition, the distorted waves can be
written as

\begin{eqnarray}
[\Psi^{(-) \,\Lambda N}_{{\vec p_F},S_F M_{S_F}} ({\vec r}\,)]^*
\,{\chi^\dagger}^{T}_{M_T}  
&=&
\sqrt{\frac{2}{\pi}}\, \sum_{J M} \,\sum_{L_F' S_F'} \,\sum_{L_F M_{L_F}}
(-i)^{L_F'} \,
[\psi^{(-)}_{\Lambda N}]^{*\,J}_{L_F' S_F', L_F S_F} (p_F, r)
\nonumber \\
&\times&\langle L_F M_{L_F} S_F M_{S_F} | J M \rangle 
\,Y_{L_F M_{L_F}} ({\hat p}_F)\, {{\cal J}^{\dagger}}^{J M}_{L_F' S_F'} 
({\hat r})
\label{eq:wfln2}
\end{eqnarray}
and
\begin{eqnarray}
\Psi^{(+)\,NN}_{{\vec p_I},S_I M_{S_I}} ({\vec r}\,)
\,\chi^{T}_{M_T}  
&=&
\sqrt{\frac{2}{\pi}}\, \sum_{J M}\, \sum_{L_I'}\, \sum_{L_I M_{L_I}}
\,i^{L_I'} \,
[\psi^{(+)}_{NN}]^J_{L_I' S_I, L_I S_I} (p_I, r)
\nonumber \\
& & \langle L_I M_{L_I} S_I M_{S_I} | J M \rangle 
\,Y_{L_I M_{L_I}}^{*} ({\hat p}_I) \,
{\cal J}^{J M}_{L_I' S_I} ({\hat r})
\label{eq:wfnn2}
\end{eqnarray}
where the quantities 
${\cal J}^{J M}_{LS} ({\hat r})$ are the generalized spherical harmonics.
The radial wave functions in the above equations are generated numerically
using the T-matrix constructed from the appropriate potentials and are
given as follows
\begin{eqnarray}
[\psi^{(-)}_{\Lambda N}]^{*\,J}_{L_F' S_F', L_F S_F} (p_F, r) &=& j_{L_F}(p_F) 
\delta _{L_FL_F'} \delta _{S_FS_F'} 
\nonumber \\
&+& \left. \int \frac {p'^2 dp' \langle p_F(L_F S_F) J | T_{\Lambda N \to \Lambda N} |
p'(L_F'S_F') J \rangle}{E_N(p_F)+E_{\Lambda}(p_F)-E_N(p')-E_{\Lambda}(p')+i\eta}
\, j_{L_F'}(p'r) \right.
\end{eqnarray}
\begin{eqnarray}
[\psi^{(+)}_{NN}]^{J}_{L_I' S_I, L_I S_I} (p_I, r) &=& j_{L_F}(p_F) 
\delta _{L_IL_I'} 
\nonumber \\
&+& \left. \int \frac {p'^2 dp' \langle p'(L_I' S_I) J | T_{NN \to NN} |
p_I(L_IS_I) J \rangle}{2E_N(p_I)-2E_N(p')+i\eta}
\, j_{L_I'}(p'r) \right.  .
\end{eqnarray}

\newpage

\newpage

\newpage

\newpage

\newpage

\newpage

\newpage

\newpage

\newpage

\begin{table}
\centering
\caption{Constants appearing in weak transition potential for the
different mesons.}
\vskip 0.2 in
\begin{tabular}{lcccc}
 $\mu_i$  & $K^{(i)}_{C}$ & $K^{(i)}_{SS}$ & $K^{(i)}_{T}$ &
$K^{(i)}_{PV}$
\\
\hline\hline
&   &    &       &         \\
$\pi$ & 0 & $\displaystyle\frac{B_\pi}{2 \overline M}
\displaystyle\frac{g_{\rm {\scriptscriptstyle NN} \pi}}{2M}$ &
 $\displaystyle\frac{B_\pi}{2 \overline M}
\displaystyle\frac{g_{\rm {\scriptscriptstyle NN} \pi}}{2M}$ &
$A_\pi  \displaystyle\frac{g_{\rm {\scriptscriptstyle NN} \pi}}{2M}$ \\
&   &    &       &         \\
$\eta$ & $0$ & $\displaystyle\frac{B_\eta}{2 \overline M}
\displaystyle\frac{g_{\rm {\scriptscriptstyle NN} \eta}}{2M}$ &
$\displaystyle\frac{B_\eta}{2 \overline M}
\displaystyle\frac{g_{\rm {\scriptscriptstyle NN} \eta}}{2M}$ &
$A_\eta  \displaystyle\frac{g_{\rm {\scriptscriptstyle NN} \eta}}{2M}$ \\
&    &     &       &         \\
K & $0$ & $\displaystyle\frac{1}{2M} \displaystyle\frac{g_{\rm
 {\scriptscriptstyle \Lambda N K}}}
{2 \overline M}$ &
$\displaystyle\frac{1}{2 M}
\displaystyle\frac{g_{\rm \scriptscriptstyle{\Lambda N K}}}{2 \overline M}$
& $\displaystyle\frac{g_{\rm \scriptscriptstyle{\Lambda N K}}}{2 M}$
\\
&    &     &       &         \\
$\rho$ & $g^{\rm {\scriptscriptstyle V}}_{\rm {\scriptscriptstyle NN} \rho}
 \alpha_\rho$ &
$2\displaystyle\frac{\alpha_\rho + \beta_\rho}{2 \overline M}
\displaystyle\frac{g^{\rm {\scriptscriptstyle V}}_{\rm {\scriptscriptstyle NN}
 \rho} +
g^{\rm {\scriptscriptstyle T}}_{\rm {\scriptscriptstyle NN} \rho}} {2M}$ &
$ - \displaystyle\frac{\alpha_\rho + \beta_\rho}{2 \overline M}
\displaystyle\frac{ g^{\rm {\scriptscriptstyle V}}_{\rm {\scriptscriptstyle NN}
 \rho} +
g^{\rm {\scriptscriptstyle T}}_{\rm {\scriptscriptstyle NN} \rho}} {2M}$ &
$ - \varepsilon_\rho
\displaystyle\frac{ g^{\rm {\scriptscriptstyle V}}_{\rm {\scriptscriptstyle NN}
 \rho} +
g^{\rm {\scriptscriptstyle T}}_{\rm {\scriptscriptstyle NN} \rho}} {2M}$ \\
&    &     &       &         \\
$\omega$ & $g^{\rm {\scriptscriptstyle V}}_{\rm {\scriptscriptstyle NN} \omega}
 \alpha_\omega$ &
$2\displaystyle\frac{\alpha_\omega + \beta_\omega}{2 \overline M}
\displaystyle\frac{g^{\rm {\scriptscriptstyle V}}_{\rm {\scriptscriptstyle NN}
 \omega} +
g^{\rm {\scriptscriptstyle T}}_{\rm {\scriptscriptstyle NN} \omega}} {2M} $ &
$ - \displaystyle\frac{\alpha_\omega + \beta_\omega}{2 \overline M}$
$\displaystyle\frac{ g^{\rm {\scriptscriptstyle V}}_{\rm {\scriptscriptstyle NN}
 \omega} +
g^{\rm {\scriptscriptstyle T}}_{\rm {\scriptscriptstyle NN} \omega}} {2M} $ &
$ - \varepsilon_\omega
\displaystyle\frac{ g^{\rm {\scriptscriptstyle V}}_{\rm {\scriptscriptstyle NN}
 \omega} +
g^{\rm {\scriptscriptstyle T}}_{\rm {\scriptscriptstyle NN} \omega}} {2M}$ \\
&    &     &       &         \\
K$^*$ & $g^{\rm {\scriptscriptstyle V}}_{\rm {\scriptscriptstyle
\Lambda N K^*}} $ &
$2 \displaystyle\frac{1}{2M} \displaystyle\frac{g^{\rm
 \scriptscriptstyle{V}}_{\rm {\scriptscriptstyle \Lambda N K^*}} +
g^{\rm \scriptscriptstyle{T}}_{\rm {\scriptscriptstyle \Lambda N K^*}}} {2
 \overline M}$  &
$ - \displaystyle\frac{1}{2 M}
\displaystyle\frac{ g^{\rm {\scriptscriptstyle V}}_{\rm {\scriptscriptstyle
 \Lambda N K^*}} +
g^{\rm \scriptscriptstyle{T}}_{\rm {\scriptscriptstyle \Lambda N K^*}}} {2
 \overline M}$ &
$- \displaystyle\frac{ g^{\rm \scriptscriptstyle{V}}_{\rm {\scriptscriptstyle
 \Lambda N K^*}} +
g^{\rm \scriptscriptstyle{T}}_{\rm {\scriptscriptstyle \Lambda N K^*}}} {2 M}$
 \\
&    &     &       &         \\
\end{tabular}
\label{tab:ctspot}
\end{table}

\end{document}